\documentstyle[preprint,revtex]{aps}
\begin{document}
\draft
\begin{title}
Hyperfine Populations Prior to Muon Capture.
\end{title}
\author{J.G.Congleton}
\begin{instit}
Instituut Voor Theoretische Fysica, Universiteit te Utrecht, \\
Postbus 80.006, 3508 TA Utrecht, The Netherlands.
\end{instit}
\begin{abstract}
It is shown that the 1S level hyperfine populations prior to muon
capture will be statistical when either target or beam are unpolarised
independent of the atomic level at which the hyperfine interaction
becomes appreciable. This assertion holds in the absence of magnetic
transitions during the cascade and is true because of minimal
polarisation after atomic capture and selective feeding during the
cascade.
\end{abstract}

The muon capture rate by a nucleus from the 1S atomic level, depends
on the populations of the hyperfine states $f,f_z$. The isotropic rate
depends only on the total populations of the two hyperfine levels
$f^{\pm}=i\pm\frac{1}{2}$, where $i$ is the total angular momentum of
the nucleus and is non-zero. In this letter, it is shown that these
populations are statistical independent of the atomic level at which
the hyperfine interaction becomes appreciable, provided at least one
of the target nucleus or the muon beam is unpolarised. This assertion
holds in the absence of magnetic transitions during the atomic
cascade.

Mukhopadhyay \cite{Muk} has shown that the populations are statistical
when the hyperfine interaction acts only at the 1S state and at least
one of the target nucleus or beam is unpolarised.  Measurement of
residual muon polarisations in light nuclei \cite{Fav} show that the
hyperfine interaction must be active at atomic states higher than
$n=1$. The question arises as to whether the populations remain
statistical when the hyperfine interaction acts before the muon
reaches the 1S state.

This question is of current interest due to the experimental proposal
\cite{Deu} to measure the statistical capture rate by \mbox{$^3$He}
to a precision of 1\% at PSI. A deviation of 2.5\% from statisticity
in the hyperfine populations would change the measured rate by 1\%
\cite{Jim} and so it is important to establish that the populations
are indeed statistical.

The population of the atomic states will be described by statistical
tensors as used by Nagamine and Yamazaki \cite{Nag} and Kuno, Nagamine
and Yamazaki \cite{Kuno} in their study of polarised muonic atoms.
The statistical tensor $B_k(j)$, $k=0,1\ldots 2j$ is proportional to
the rank $k$ polarisation of the level $\mid \! j \, \rangle$ and is
defined below.  With this definition, $B_0(j)$ equals the population
of the level $\mid \! j \, \rangle$.
\FL \begin{equation}
B_k(j) = \sqrt{2j+1} \sum_m (-1)^{j-m} P_m
\langle k 0 \! \mid j j m -\!m \rangle
\end{equation}
where $P_m$ is the population of the state $\mid \! j,m \, \rangle$.
The conventions for the angular momentum algebra follow Brink and
Satchler \cite{Bri} throughout this letter. At atomic capture the
atomic orbitals are filled without prejudice to $m_l$. This
corresponds to the muons having negligible angular correlation with
the beam direction \cite{Rose}.  After some fast internal Auger
transitions the spin-orbit interaction splits terms according to $j$
and the statistical tensors\ for the level $n,l,j$ are,
\FL \begin{equation} \label{capture}
B_k(n,l,j) = \left [ \frac{(2j+1)^3}{(2l+1)(2s+1)} \right ]
^\frac{1}{2}
\sum_{k_1,k_2} B_{k_1}(l) B_{k_2}(s)
\, \langle k0 \!\mid k_1k_200 \, \rangle
\left \{\begin{array}{ccc} l&s&j \\ l&s&j \\ k_1&k_2&k \end{array}
\right \}.
\end{equation}

Only $B_0(l)$ is non-zero and using the triangular selection rule
imposed by the Clebsch-Gordan coefficient we see that only
$B_0(n,l,j)$ and $B_1(n,l,j)$ are non-zero. This `minimal
polarisation' feature is a direct result of the isotropy of the muons
just before atomic capture and the fact that muons are spin
$\frac{1}{2}$ particles. $B_1(n,l,j)$ is proportional to $P_{\mu}$,
the polarisation of the muon before atomic capture.

There follows an electromagnetic cascade whereby the muonic atom
de-excites from $n \approx 14 $ to the 1S level.  For electric
transitions of multipolarity $L$, the new statistical tensors due to
the $n,l,j \rightarrow n',l',j'$ transition are,
\begin{eqnarray} \label{trans}
B_k(n',l',j') &=& (2l+1)(2j'+1) \,
W(jj'll';L\raisebox{.4ex}{$\scriptstyle\frac{1}{2}\displaystyle$})^2
\, u_k(jLj') \, B_k(n,l,j) \\
u_k(jLj') &=& (-1)^{k+L-j-j'} [(2j+1)(2j'+1)]^{\frac{1}{2}}
W(jjj'j';kL) .\end{eqnarray}
Thus, the statistical tensor of rank $k$ is fed only by the rank $k$
statistical tensors for higher levels. This feature may be termed
`selective feeding' since the tensor $B_k$ feeds other tensors
according to the selection rule $\Delta k =0$.  It follows that $B_0$
and $B_1$ will be the only non-zero statistical tensors\ during the
cascade.

At the level where the hyperfine splitting becomes larger than the
natural width, the statistical tensors\ for the states $\mid \!
n,l,j,f \, \rangle$ are,
\FL \begin{equation} \label{hyp}
B_k(n,l,j,f) = \left [ \frac{(2f+1)^3}{(2i+1)(2j+1)} \right ]
^\frac{1}{2}
\sum_{k_1,k_2}  B_{k_1}(i) B_{k_2}(n,l,j)
\,  \langle k0 \! \mid k_1k_200 \, \rangle
\left \{\begin{array}{ccc} i&j&f \\ i&j&f \\ k_1&k_2&k \end{array}
\right \},
\end{equation}
where $B_{k_1}(i)$ is the statistical tensor for the nucleus.  By
observing the selection rule for $k$, $k_1$ and $k_2$, the total
populations for the hyperfine states must have the following form,
\begin{equation} \label{par}
B_0(n,l,j,f) = \alpha + \beta P_{\mu} P_i
\end{equation}
where $\alpha$ and $\beta$ are constants peculiar to the level
\mbox{$\mid \! n,l,j,f \, \rangle$} and $P_i$ is the vector
polarisation of the nucleus.  Higher rank polarisations of the nucleus
cannot contribute to $B_0(n,l,j,f)$ since the rank of the total
angular momentum polarisation is no higher than 1.

Allowing the hyperfine levels to decay only via electric transitions
we have for the transition $n,l,j,f \rightarrow n',l',j',f'$,
\begin{eqnarray} \label{transf}
B_k(n',l',j',f') &=& (2l+1)(2j'+1) \,
W(jj'll';L\raisebox{.4ex}{$\scriptstyle\frac{1}{2}\displaystyle$})^2
\times (2j+1)(2f'+1)
W(ff'jj';L\raisebox{.4ex}{$\scriptstyle\frac{1}{2}\displaystyle$})^2
\nonumber \\
&& \, \times u_k(fLf') \,B_k(n,l,j,f)
\end{eqnarray}
which has the same `selective feeding' property as equation
(\ref{trans}).  Using this fact and equation (\ref{par}) the 1S
hyperfine populations can be parameterised by,
\begin{equation}
B_0(1,0,\raisebox{.4ex}{$\scriptstyle\frac{1}{2}\displaystyle$},f) =
\gamma + \delta P_{\mu} P_i .\end{equation} The parameter
$\gamma$ may be found either by demanding that the populations are
statistical when $P_{\mu}=P_i=0$ or by directly calculating it using
equations (\ref{capture}) (\ref{trans}) (\ref{hyp}) and
(\ref{transf}).
\begin{equation}
\gamma = \frac{2f+1}{2(2i+1)} \end{equation}

Thus, if either $P_{\mu}=0$ or $P_i=0$, the total populations of the
1S hyperfine levels are statistical in the limit that no magnetic
transitions occur during the cascade. This is due to minimal
polarisation after atomic capture and selective feeding during the
cascade.

\acknowledgments

This research was supported by the British Royal Society under the
auspices of the European Science Exchange Programme.

\end{document}